\documentclass[iop]{emulateapj}
\usepackage{lipsum}
\usepackage{epstopdf}
\usepackage{natbib}
\usepackage{courier}
\usepackage{hhline}
\usepackage{booktabs}
\usepackage{multirow}

\shorttitle{PDR Modeling in the Taurus Molecular Cloud}
\shortauthors{Orr, Pineda, \& Goldsmith}

\begin{document}

\title{Photon-Dominated Region Modeling of the [C\,{\sc i}],[C\,{\sc ii}], and CO Line\\ Emission
    from a Boundary in the Taurus Molecular Cloud}


\author{Matthew E. Orr\altaffilmark{1,2}, Jorge L. Pineda\altaffilmark{2}, and Paul F. Goldsmith \altaffilmark{2}}
\affil{$^1$Physics \& Astronomy Department, University of Southern California,
    Los Angeles, CA 90089\\$^2$Jet Propulsion Laboratory, California Institute of Technology, 4800 Oak Grove Drive, Pasadena, CA 91109-8099, USA}



\begin{abstract}
We present [C\,{\sc i}] and [C\,{\sc ii}] observations of a linear edge region in the Taurus molecular cloud, and model this region as a cylindrically symmetric PDR exposed to a low-intensity UV radiation field.  The sharp, long profile of the linear edge makes it an ideal case to test PDR models and determine cloud parameters.  We compare observations of the [C\,{\sc i}], $^3$P$_{1}\rightarrow^3$P$_{0}$ (492 GHz), [C\,{\sc i}] $^3$P$_2 \rightarrow ^3$P$_1$ (809 GHz), and [C\,{\sc ii}] $^2$P$_{3/2}\rightarrow^2$P$_{1/2}$ (1900 GHz) transitions, as well as the lowest rotational transitions of $^{12}$CO and $^{13}$CO, with line intensities produced by the RATRAN radiative transfer code from the results of the Meudon PDR code.  We constrain the density structure of the cloud by fitting a cylindrical density function to visual extinction data.  We study the effects of variation of the FUV field, $^{12}$C/$^{13}$C isotopic abundance ratio, sulfur depletion, cosmic ray ionization rate, and inclination of the filament relative to the sky-plane on the chemical network of the PDR model and resulting line emission.  We also consider the role of suprathermal chemistry and density inhomogeneities.  We find good agreement between the model and observations, and that the integrated line intensities can be explained by a PDR model with an external FUV field of 0.05 $G_0$, a low ratio of $^{12}$C to $^{13}$C $\sim 43$, a highly depleted sulfur abundance (by a factor of at least 50), a cosmic ray ionization rate $(3-6) \times 10^{-17}$ s$^{-1}$, and without significant effects from inclination, clumping or suprathermal chemistry.

\end{abstract}



\section{Introduction}
\begin{centering}
\begin{figure*}[!t]
\vspace{-10mm}
\centering
\includegraphics[width=\textwidth]{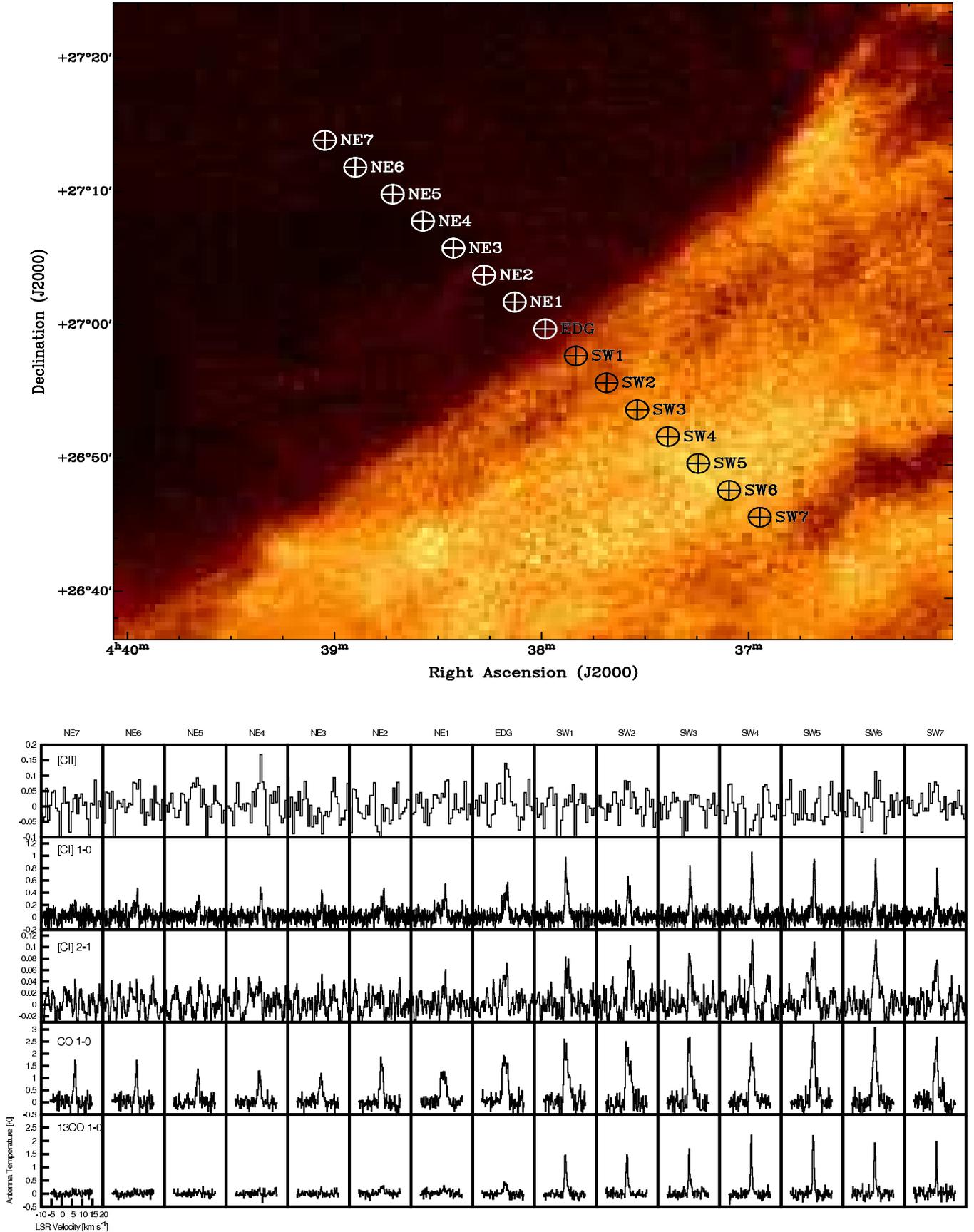}
\caption{Spectra of [C\,{\sc ii}], [C\,{\sc i}], $^{12}$CO and $^{13}$CO at the selected points, labeled at top of the lower panel, across the linear edge region in Taurus.  The top panel is a map of the region in $^{13}$CO peak intensity, with observation points indicated.  A marked jump in antenna temperature for molecular and atomic lines is observed as one moves into the cloud across the edge (denoted EDG).  Possible detections of [C\,{\sc ii}] are at locations NE4 and EDG.}
\label{fig:panel}
\end{figure*}
\end{centering}
New stars in our galaxy are born deep in the cold, dense cores of molecular clouds.  The formation of these clouds, and their evolution, are of great interest to our understanding of star formation.  The heating and chemical processes of the boundaries of these clouds, which mark the transition from a region of primarily atomic to primarily molecular species, are dominated by far-ultraviolet (FUV; 6 eV $\lesssim h\nu \lesssim$  13.6 eV) photon-driven processes.  For this reason, these regions, whose characteristics are dependent on the FUV flux, are known as photon dominated regions (PDRs) \citep[][and references therein]{hollenbach1999,snow2006}.

Across these boundary regions, the fractional abundances of species present vary considerably.  In general the fractional abundances of molecules are highest in the dense interiors of clouds, and drop in the less well-shielded outer envelopes of PDRs, while the abundances of atomic and ionized species often peak in the transition regions and in the outer envelopes of PDRs.  The thermal balance in these regions is primarily driven by photoelectric effect heating and line emission cooling; a discussion of the physical processes involved in the thermal balance is found in \citet{hollenbach1999}.  The chemistry is ultimately determined by the energy input to the region in the form of FUV photons and cosmic rays.  Additionally, the relative abundances of elements, electron density, and dust grain size and distribution affect the chemical networks of PDRs. Due to this complexity, over the past several decades PDRs have been extensively studied, and it is thought that the physical and chemical processes that occur within them are relatively well known.

The interdependence of the heating, cooling, and the chemical balance of PDRs make for solving their equations of state rather difficult.  In recent years, a number of numerical models have been developed which seek to solve for the detailed chemical and physical structure of PDRs for a number of species of interest, such as the Leiden code \citep{black1987}, CLOUDY \citep{ferland2013}, the Meudon PDR code \citep{lepetit2006}, and KOSMA\textendash$\tau$ \citep{rollig2006}.  A comparison of a number of these codes' efficacy, and of differences among them has been published \citep{rollig2007}. 

Most studies of PDRs have focused on bright, massive star formation regions, with ambient far-ultraviolet flux $G_0\gg1$ \citep[in units of the average interstellar flux of $1.6 \times 10^{-3}$ erg cm$^{-2}$ s$^{-1}$;][]{habing1968}.  Relatively little has been done on low $G_0$ environments ($G_0<1$), which are rather quiescent, with low dust and gas temperatures.  Previous work by \citet{pineda2007} investigated a PDR in the envelope of the B68 cloud, with a constrained density profile and $G_0 \sim 0.59$, and found difficulty matching $^{13}$CO line emission observations with their models to within a factor of 2.  This has left open the question of whether or not computational PDR models satisfactorily treat extremely low FUV fluxes and their consequences.  

In part for this reason, we have observed a boundary in the Taurus molecular cloud \citep[at a distance of 140 pc;][]{kenyon1994}, which has an especially favorable PDR boundary for modeling due to its sharp well-defined edge and relatively simple geometry.  We used the Herschel Space Observatory (HSO), as well as the Five College Radio Astronomy Observatory (FCRAO) for our observations. Previous studies suggest that this region has a relatively low FUV field \citep{flagey2009, pineda2010}, and little foreground or background visual extinction \citep{padoan2002}, making it favorable for the comparison of observations with physical models.  In addition, this region of Taurus has been investigated by \citet{goldsmith2010}, observing H$_2$ emission in this boundary region.  They found warm molecular hydrogen with a very small column density, $(1 - 5) \times$10$^{18}$ cm$^{-2}$, which is insufficient to significantly perturb any of the tracers studied here. Goldsmith et al. concluded that additional heating beyond that provided by the interstellar radiation field must be present to explain the high excitation temperatures observed.  Together, these previous studies have made this boundary region in Taurus an interesting testbed for studying the interface between molecular clouds and their surroundings.

In this paper, we present observations of the [C\,{\sc i}] $^3$P$_1\rightarrow^3$P$_0$ and $^3$P$_2\rightarrow^3$P$_1$, and [C\,{\sc ii}] $^2$P$_{3/2}\rightarrow^2$P$_{1/2}$ transitions of a boundary in the Taurus molecular cloud.  We complement these data with observations of the $^{12}$CO and $^{13}$CO $J =1\rightarrow0$ transitions of the region presented in \citet{narayanan2008} and \citet{goldsmith2008}.  We also model the photochemistry and temperature profile of this region using the Meudon PDR code.  We use the RATRAN radiative transfer code to compare the model's line emission predictions with observations.  Finally, we discuss the implication of our model's best-fit parameters, and possible further work.

\section{Observations}\label{obs}

\subsection{[C\,{\sc i}] \&   [C\,{\sc ii}] Observations}\label{c_obs}
We observed the Taurus molecular cloud in the [C\,{\sc i}] $^3$P$_1 \rightarrow ^3$P$_0$,  [C\,{\sc i}] $^3$P$_2 \rightarrow ^3$P$_1$, and   [C\,{\sc ii}] $^2$P$_{3/2} \rightarrow ^2$P$_{1/2}$ fine structure lines at 492.1607 GHz, 809.3420 GHz, and 1900.5469 GHz (rest frequency), respectively, with the HIFI \citep{deGraauw2010} instrument aboard the \emph{Herschel Space Observatory}\footnote{The Herschel Science Archive project ID was OT1\_pgolds01\_5} \citep{pilbratt2010}.  

Spectra were taken at 15 positions across the linear edge boundary, as seen in Figure \ref{fig:panel}.  Table \ref{tab:intens} gives the velocity integrated intensities.  These positions were spaced by $\simeq 3'$, giving a good sampling across the edge region.

The [C\,{\sc i}] lines were observed with the HIFI Band 1 and 3 receivers.  The [C\,{\sc ii}] line was observed with the HIFI Band 7b receiver, which uses Hot Electron Bolometer (HEB) mixers.  Both the [C\,{\sc i}] and [C\,{\sc ii}] observations made use of the LoadChop with reference observing mode.  This mode allows use of an internal cold calibration source as a comparison load to account for short term changes in the instrument's behavior.  As there exists a difference in the optical path between the observed source and the internal comparison load, a residual standing wave structure is produced, which is then removed by subtraction of a sky reference position.  Our reference position for the [C\,{\sc i}] and [C\,{\sc ii}] lines was $\alpha = 4^h 40^m 13.464^s$, $\delta = 27^\circ 21' 33''$ (J2000).  It is known that data from the Band 7b receiver has prominent electrical standing waves, which are generated between the HEB mixer and the first low noise amplifier.  Since the spectral feature in the [C\,{\sc ii}] emission in Taurus was expected to be a single, narrow component, we removed the standing waves by fitting a combination of sinusoidal functions, using the \texttt{FitHIFIFringe()} procedure in \texttt{HIPE} \citep{ott2006}.  The resulting [C\,{\sc ii}] spectra, as well as the [C\,{\sc i}] spectra, were then exported to be analyzed by the  \texttt{CLASS90}\footnote{http://www.iram.fr/IRAMFR/GILDAS/} data analysis software suite, which we used to combine the two linear polarizations and fit polynomial baselines (normally of order 3).

The angular resolution for the [C\,{\sc i}] lines was 44$''$ at 492 GHz, 26.5$''$ at 809 GHz, and 12$''$ for the 1900 GHz [C\,{\sc ii}] transition.  We applied main-beam efficiencies of 0.79, 0.78, and 0.72, respectively, to convert these data from an antenna temperature to a main-beam temperature scale \citep{roelfsema2012}.                                                                                                                                                                                                                                                                                                                   

The [C\,{\sc i}] data were taken with the high resolution spectrometer (HRS) for the 492 GHz transition, and the wide band spectrometer (WBS) for the 809 GHz transition.  These have channel widths of 128 KHz and 1.1 MHz, or 0.078 km s$^{-1}$ at 492 GHz and 0.41 km s$^{-1}$ at 809 GHz \citep{roelfsema2012}.  The data were not further smoothed in velocity.  The average rms noise was 0.07 K, and 0.02 K for the two [C\,{\sc i}] lines with the channel widths given above. 

The [C\,{\sc ii}] observations used the WBS, with a channel width of 1.1 MHz (0.17 km s$^{-1}$ at 1900.5 GHz).  These data were smoothed in velocity to a resolution of 0.96 km\,s$^{-1}$, with an average rms noise of 0.04 K.  Except for two possible detections, at the edge (EDG) and 11.3$'$ (NE4) offset from the linear edge, [C\,{\sc ii}] was not detected in at any individual position in this region of the Taurus molecular cloud.
\begin{centering}
\begin{table}[!t]
\centering
\caption{Integrated Intensities of Observed Transitions across the Linear Edge Region in Taurus}
\label{tab:intens}
\begin{tabular}{cccccc}
\hline\\\vspace{-6.3mm}\\\hline\\
Position &\multicolumn{1}{c}{Offset} & \multicolumn{4}{c}{Intensity (K km s$^{-1}$)} \\\cline{3-6}\\
ID &   ( $'$ ) &  [C\,{\sc i}]   &  [C\,{\sc i}] &  $^{12}$CO & $^{13}$CO \\
 &  &492 GHz&809 GHz&115 GHz&110 GHz\\\hline
  
NE7  & 20.1 &  0.50 &  0.02  & 5.33  &  -0.35\\
NE6  & 17.3 &  0.84 &  0.08  & 4.03 &  -0.06\\
NE5  & 14.1 &  0.46 &-0.02  & 4.33 &  0.68\\
NE4  & 11.3 &  0.71 & 0.14  & 3.81&  0.45\\
NE3  &   8.5 &  0.46 & 0.06  & 3.73 &  -0.19\\
NE2  &   5.7 &  0.82 & 0.04  & 6.96&  1.48\\
NE1  &   2.8 &  1.01 & 0.08  & 7.32&  1.23\\
EDG  &   0.0 &  1.26 & 0.19  &12.63&  1.52\\
SW1  &  -2.8 &  1.62 & 0.16  &12.96&  3.36\\
SW2  &  -5.7 &  1.03 & 0.19  &13.00&  3.28\\
SW3  &  -8.5 &  1.10 & 0.18  &11.16&  3.22\\
SW4  &-11.3 &  1.50 & 0.23  &10.64&  4.14\\
SW5  &-14.1 &  1.74 & 0.27  &12.96&  4.58\\
SW6  &-17.0 &  1.20 & 0.28  &13.36&  3.89\\
SW7  &-19.8 &  0.72 & 0.22  &  9.59&  2.59   \\\hline\\
   $\sigma^a$&  & 0.077&  0.033  &  0.53  &  0.30   \\\hline\\
\multicolumn{6}{p{.450\textwidth}}{ $^a$ Typical root-mean-square noise in a single channel.}\\
\multicolumn{6}{p{.450\textwidth}}{\textbf{Note.} [C\,{\sc ii}] 1.9 THz observations were not included in this table due to absence of definitive detection.  See spectra in Figure \ref{fig:panel}.}\\\vspace{2 mm}
\end{tabular}
\end{table}
\end{centering}

\subsection{$^{12}$CO \& $^{13}$CO Observations}\label{co_obs}
The $^{12}$CO and $^{13}$CO observations were taken using the 13.7 m Five College Radio Astronomy Observatory (FCRAO) Telescope.  The observations took advantage of the 32 pixel SEQUOIA focal plane array (a 16 pixel single polarization version of this array is described in \citealt{erickson1999}).  The SEQUOIA instrument simultaneously observed the $J = 1 \rightarrow 0$ transition of $^{12}$CO and $^{13}$CO, at 115.2712 GHz and 110.2014 GHz (rest frequency), respectively.  The receiver array consists of two sets of 4$\times$4 pixel arrays, in two orthogonal linear polarizations.  Details of the data acquisition, reduction, and calibration procedures are given by \citet{narayanan2008} and \citet{goldsmith2008}.

The main beam of the antenna pattern had a full-width to half-maximum (FWHM) beam-width of 45$''$ for $^{12}$CO, and 47$''$ for $^{13}$CO.  The data have a mean rms antenna temperature of 0.28 K and 0.125 K, in channels of 0.26 km s$^{-1}$ and 0.27 km s$^{-1}$ width for $^{12}$CO and $^{13}$CO, respectively.

\subsection{Visual Extinction Data}\label{Av_obs}
We used the visual extinction data from \citet{pineda2010} to fit the density structure for our model cloud. Details of the observations, techniques of data reduction, and their analysis of the data can be found in that paper.  This data provided us with a reliable measure of total H$_2$ column density along the line of sight through our region of interest, thereby constraining our PDR model.

\section{Analysis \& Modeling}\label{anal}
We assumed a cylindrical geometry to model the linear edge region of the cloud. We then employed a one-dimensional, plane parallel slab PDR model developed by \citet{lepetit2006}, which uses an isotropic FUV field.  To model cuts through the cylindrical edge geometry, a slab model was not adequate.  In order to match our results to our sky-plane data, we mapped the 1-d solutions for gas and dust temperature and chemical abundances to a cylindrical geometry.  We transferred the results of our ``cylindericalized" Meudon PDR model to a spherical 3-d radiative transfer model, RATRAN, developed by \citet{hogerheijde2000}, and were thus able to generate a sky-plane image of our linear edge region model for the radiative transitions of the species of interest.  Assuming the radius of curvature of that spherical RATRAN model to be very large, we take the line intensity results of a plane cut through the center of the spherical model as approximating the results of a plane cut through the cylindrical model perpendicular to its axis.  We iterated over a wide range of physical parameters in order to match the data most accurately.
\subsection{Linear Edge Density Profile}
Initially, we model the linear edge as a cylinder with a radius-dependent H$_2$ volume density structure of the following form \citep{king1962},
\begin{equation}
   n_{\rm H_2}(r) = \left\{
     \begin{array}{lr}
       n_ca^2/(r^2 + a^2) & : r \leq R\\
       0 & : r > R 
     \end{array} .
   \right.
\label{eq:king}
\end{equation} 

This functional form for the density structure was proposed by \citet{dapp2009} for dense H$_2$ cores, and subsequently used in analysis of the Taurus region by \citet{pineda2010}.

This profile has a flat high-density center, which then transitions to a region of power-law decay, and finally, at a truncating radius, is set to zero.  It assumes only three parameters, the central core density $n_c$, a parameter characterizing the width of the central core $a$, and the truncating radius $R$.  

Integrating the density profile along lines of sight through the cylindrical model results in the following relation for H$_2$ column density $N_{\rm H_2}(x)$, centered on the core of the cylindrical model, as a function of linear offset $x$,
\begin{equation}\label{colds}
N_{\rm H_2}(x)=\frac{2an_c}{\sqrt{1+(x/a)^2}}\arctan\left(\sqrt{\frac{R^2-x^2}{a^2+x^2}}\right).
\end{equation}
\begin{center}
\begin{figure}[!t]
\centering
\includegraphics[width=0.5\textwidth]{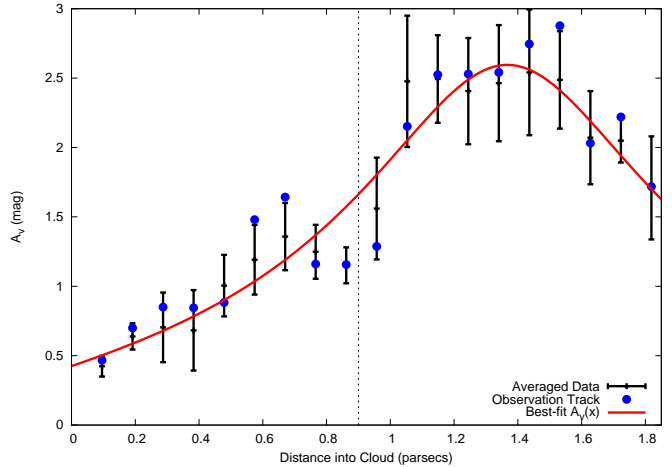}
\caption{Plot of the best-fit radial density profile's resulting visual extinction ($A_{\rm v}$) profile and observed $A_{\rm v}$ data from \citet{pineda2010}, across a cut through the linear edge region.  The dashed vertical line indicates the position of the EDG observation location along the cut (see Figure \ref{fig:panel}).  Observed $A_{\rm v}$ data were averaged with nearby parallel tracks in order to achieve a better fit.  The radial density profile was fit to this smoothed data.  $A_{\rm v}$ data from the observational track of our spectra (see Figure \ref{fig:panel}) are also shown; note that these data are taken along the same cut across the linear edge as spectral observations, but not at the same points. One standard deviation error bars are included for the smoothed data.\\}
\label{fig:kingfit}
\end{figure}
\end{center}
\vspace{-4mm}

Implicitly, we also assume that the cylinder lies in the plane of the sky.  If this were not the case, the central density $n_c$ of the filament would be replaced by $n_c \sec(\theta)$, accounting for the inclination of the filament with respect to the sky-plane by an angle $\theta$.  This would result in lower values for $n_c$.  Inclination of the cylinder relative to the sky-plane did not appear to be supported by observations, and effects of inclination of the filament on the line integrated intensities will be discussed separately later in Section \ref{inc}.  All values presented hereafter are for the filament lying in the sky-plane ($\theta = 0$).

Comparing the relation (Eq.\,\ref{colds}) with visual extinction data for our region of interest, taken from \citet{pineda2010}, we were able to fit the profile closely to observations, as seen in Figure \ref{fig:kingfit}.  Observed $A_{\rm v}$ data were averaged with nearby cuts through the linear edge, parallel to that of the observations, in order to reduce noise and obtain a more representative profile.  We converted the visual extinction to H$_2$ column density using $N($H$_2)/A_{\rm V}=9.4\times10^{20}$ cm$^{-2}$ mag$^{-1}$, obtained by assuming that the hydrogen is predominately in the molecular form, allowing us to express the color excess observed by \citet{bohlin1978} as $N($H$_2)/E_{\rm B-V}=2.9\times10^{21}$ cm$^{-2}$ mag$^{-1}$, and combining this with the ratio of total to selective extinction $R_{\rm V}= A_{\rm V}/E_{\rm B-V} \simeq 3.1$ \citep{whittet2003}.  The fitted values for the density profile parameters are $n_c = 626$ cm$^{-3}$, $a =0.457$ pc, and $R = 1.80$ pc.  The axis of the cylindrical distribution is centered $11.5'$ to the southwest of the EDG position.

\subsection{PDR Modeling}
The Meudon PDR code requires many physical parameters in addition to a linear density profile to model the PDR.  The most important for our modeling are the strength of the FUV field $G_0$, the initial relative fractional elemental abundances, and the cosmic ray ionization rate $\zeta$.

Using the aforementioned parameters, we were able to generate 1-d numerical solutions for the temperature and relative fractional abundances of species, as a function of linear depth into the PDR.  Assuming the radius of curvature of the actual region to be sufficiently large, the variation with distance from the slab surface approximates the variation with radius from the cylindrical surface of the filament, and we thus assume that these results approximate the radial solution for the cylindrical model for the Taurus edge region.

The strength of the local FUV field determines many attributes of a PDR.  The flux level dictates dust and gas temperatures from the surface of the cloud inwards, and drives the majority of the chemistry in the clouds, with exception of the processes occurring deep within dark cores, where cosmic ray ionization gains greater importance as the FUV flux is attenuated.  Previous studies have suggested that a low FUV field exists, with a $G_0$ between 0.059 and 0.59 \citep[compared to $G_0 \sim 1.7$ for the local interstellar radiation field;][]{draine1978} in the Taurus region \citep{pineda2010}.  However, we note that recent work suggests that a higher $G_0$ of 4.25 is necessary to explain hydrogen volume densities in the Taurus region \citep{heiner2013}.  We restricted our PDR models to low FUV fluxes, adopting values of $G_0$ between 0.05 and 0.25, as suggested by \citet{pineda2010}.

Coupled with the FUV flux, the initial relative fractional elemental abundances in the PDR model are very important in determining the chemical equilibrium as a function of depth into the PDR.  The chemical network of the \citet{lepetit2006} model includes H, He, C, O, N, S, and Fe.  $^{13}$C and $^{18}$O are both included (the second-most abundant isotopes of C and O), as well as their isotopically-selective reaction pathways.  In our models, we adopted values for the initial relative fractional chemical abundances given in Table \ref{tab:chem}.

\begin{centering}
\begin{table}[h]
\centering
\caption{Initial Fraction Elemental Abundances in PDR Models$^a$}
\label{tab:chem}
\begin{tabular}{cc}
\hline\\\vspace{-6.3mm}\\\hline\\
Species & Initial Abundance \\\cline{1-2}\vspace{-2mm} \\
He & 0.1\\
C  & $1.32 \times 10^{-4}$ \\
O   & $3.19 \times 10^{-4}$ \\
N  & $7.50 \times 10^{-5}$ \\
$^{13}$C    & $(1.71  - 3.07) \times 10^{-6}$ \\
S   & $(0.186 - 1.86) \times 10^{-6}$ \\
$^{18}$O   & $6.38 \times 10^{-7}$ \\
Fe & $1.50 \times 10^{-8}$ \\\hline\\
\multicolumn{2}{p{.25\textwidth}}{$^a$ Abundance relative to total number density (i.e. H nuclei = 0.9).}\\\vspace{2 mm}
\end{tabular}
\end{table}
\end{centering}
As some uncertainty exists as to the correct value for the $^{12}$C to $^{13}$C abundance ratio, we varied this ratio in our simulations by holding the abundance of $^{12}$C constant, and varying that of $^{13}$C.  The isotopic ratio affects the rates of isotopically-selective processes that may enhance production of $^{13}$CO and C$^{18}$O, and is therefore important for predicting accurate line intensities \citep{bally1982}.  The literature suggested that the value of $^{12}$C/$^{13}$C is between 43 and 77 in the Taurus region.  We restricted our investigation to several values across this range (43, 57, 65, and 77), found for the local Galactic region by \citet{wilson1981}, \citet{hawkins1987}, \citet{langer1990}, and \citet{wilson1994}.  

Sulfur is an abundant metal and is easily ionized, so is often the dominant source of electrons in dense regions of the ISM.  As the fractional abundance of S$^+$ relative to C$^+$ affects rates of chemical pathways that determine equilibrium CO and C abundance values.  Previous models have found that sulfur must be substantially depleted in dense, well--shielded regions in order to get [CO/H$_2$] to approach 10$^{-4}$.  Thus, \citet{graedel1982} depleted sulfur by a factor $\simeq$ 30 in their ``low metal abundance" model, while in standard UMIST model, \citet{mcelroy2013} has a sulfur depletion factor greater than 100.  These large depletions are consistent with comparisons of the CO abundance with direct tracers of the total column density, which confirm that the abundance of CO does reach 10$^{-4}$ \citep{pineda2010}, as well as with recent chemical models \citep{visser2009}.  We thus considered initial sulfur depletion factors, $\delta$(S), in our model of 10, 20, 50 and 100 in order to investigate the effects of sulfur depletion in the observed line intensities.

The last parameter we varied was the cosmic ray ionization rate, a factor which drives heating and chemistry deep within molecular clouds where the FUV flux has been greatly attenuated.  It has been proposed by \citet{mccall2003} that high fluxes of low-energy cosmic rays are needed to explain H$_3^+$ column densities in diffuse clouds having low visual extinction $\simeq$ 0.3 mag. with cosmic ray ionization rates as high as $\zeta = 1.2 \times 10^{-15}$ s$^{-1}$.  Much lower cosmic ray ionization rates were found by \citet{goldsmith2005} whose calculations of total cosmic ray fluxes found best agreement between theoretical and observed core temperatures with a cosmic ray ionization rate of $\zeta = 5.2 \times 10^{-17}$ s$^{-1}$, and set an upper limit on the cosmic ray ionization rate of  $\zeta = 5 \times 10^{-16}$ s$^{-1}$ for thicker, denser clouds i.e. $A_{\rm V} \geq 0.3$ mag.  The high flux of low-energy cosmic rays proposed by \citet{mccall2003} may likely be absorbed in the outer envelopes of the PDR where FUV flux still dominates, and agreement may be found with studies of dark cores if this is indeed the case where there is a component of the cosmic-ray spectrum that penetrates diffuse but not dense clouds.  Additionally, \citet{doty2004} suggest that the much higher rate McCall et al. derive for IRAS 16293--2422 is actually being dominated by X-ray ionization from the central star, and that the true cosmic ray ionization rate is much lower.  As the region we investigate is relatively dense and believed to be rather quiescent, we confined $\zeta$ to the lower range of $(3-9) \times 10^{-17}$ s$^{-1}$ in our models. As will be discussed at the end of Section 4, we find that the cosmic ray ionization rate is not a critical parameter for fitting the model to the present data.

\begin{center}
\begin{figure}[!t]
\centering
\includegraphics[width=0.5\textwidth]{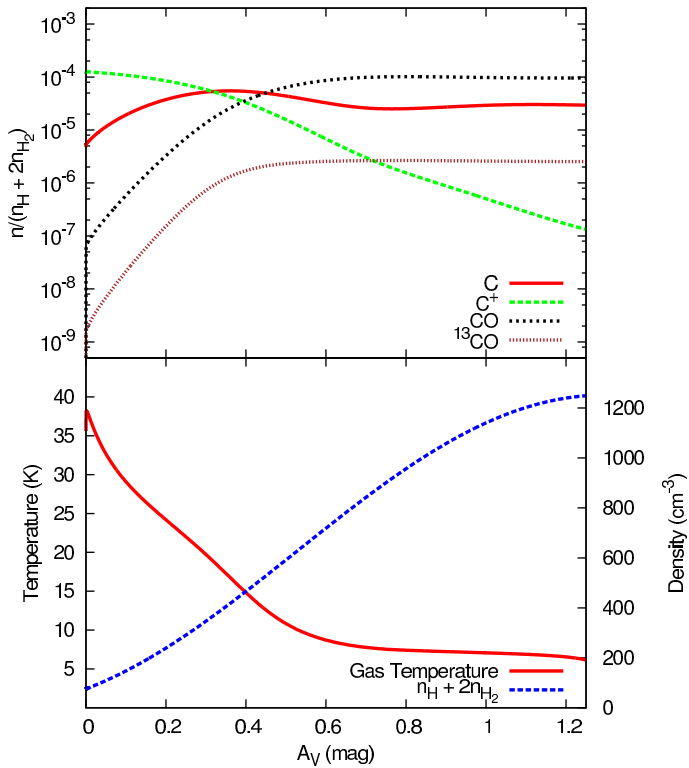}
\caption{Output from the Meudon PDR code of the best representative model of the linear edge region, with $G_0 = 0.05$, $\delta$(S)=50, $^{12}$C/$^{13}$C=43, and $\zeta=3\times10^{-17}$ s$^{-1}$.  Top Panel: Fractional abundances of observed species, relative to total hydrogen density, as a function of extinction depth into the cylindrical model.  Bottom Panel: Temperature and total hydrogen number-density profiles, as a function of visual extinction depth.\\}
\label{fig:Meudonout}
\end{figure}
\end{center}
\vspace{-10mm}
\subsection{Radiative Transfer Modeling}
We used the RATRAN code of \citet{hogerheijde2000} to calculate level populations and subsequently the line profiles for  [C\,{\sc i}] $^3$P$_1\rightarrow^3$P$_0$ and $^3$P$_2\rightarrow^3$P$_1$, [C\,{\sc ii}] $^2$P$_{3/2}\rightarrow^2$P$_{1/2}$, and $^{12}$CO and $^{13}$CO $J =1\rightarrow0$.  In RATRAN, we took the PDR model results for chemical abundance and temperature profiles to be the radial solutions of a spherical cloud with the same radial density profile used in our linear edge region.  We again assumed the radius of curvature of the cloud was sufficiently large, that our plane-parallel approximation holds.  In modeling the level populations and calculating line profiles, we adopted a FWHM line width, doppler b-parameter of $\Delta v= 1$ km s$^{-1}$, in order to have line widths consistent with those found in \citet{pineda2010} for regions having $A_{\rm V} \leq 3$ mag. in Taurus.

Having generated line profiles, we integrated these in velocity using the MIRIAD package \citep{sault1995} to compare with the observed line intensities across our linear edge region.  We took a centerline cut through our RATRAN-generated sky-plane image to compare with our observations, as we assumed the cut through the center of our spherical RATRAN model closely approximates a cut through our cylindrical model of the edge region, which has the same cross-section, both in geometry and density function.

\subsection{Comparison with Observations}

We compared the resulting model intensities directly with observations without convolving Gaussian antenna beam patterns into the images, as the FWHMs of the observing beams are much smaller ($\leq 45''$) than the size of our structure ($\sim 1^{\circ}$). 
\section{Results \& Discussion}\label{res}

\begin{center}
\begin{figure}[!t]
\centering
\includegraphics[width=0.46\textwidth]{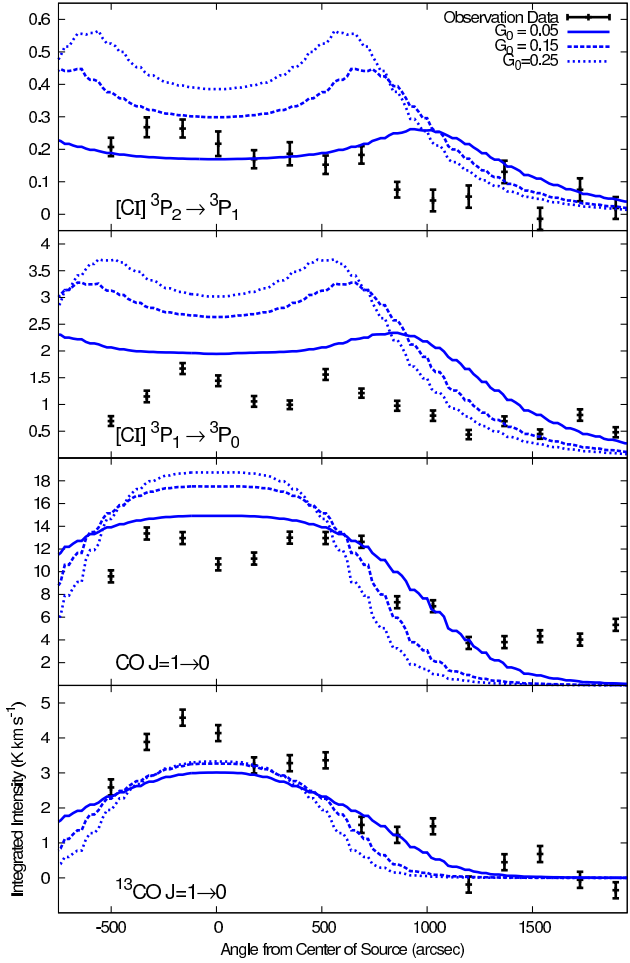}
\caption{Effect of variation of the interstellar radiation field far-ultraviolet flux (FUV) $G_0$, on [C\,{\sc i}], $^{12}$CO, and $^{13}$CO integrated line intensities, as a function of angular offset from the center of the cylindrical model.  Models compared with observational data for several values of $G_0$, with other parameters being: $\delta({\rm S}) = 50$, and $^{12}$C/$^{13}$C = 43, $\zeta=3\times10^{-17}$ s$^{-1}$.  Panels from Top to Bottom:  [C\,{\sc i}] $^3$P$_{2}\rightarrow^3$P$_{1}$ (809 GHz),  [C\,{\sc i}] $^3$P$_1 \rightarrow ^3$P$_0$ (492 GHz), $^{12}$CO $J =1\rightarrow0$ (115 GHz), and $^{13}$CO $J =1\rightarrow0$ (110 GHz) transitions, models and data.\\}
\label{fig:G0_var}
\end{figure}
\end{center}
\begin{center}
\begin{figure}[!t]
\centering
\includegraphics[width=0.46\textwidth]{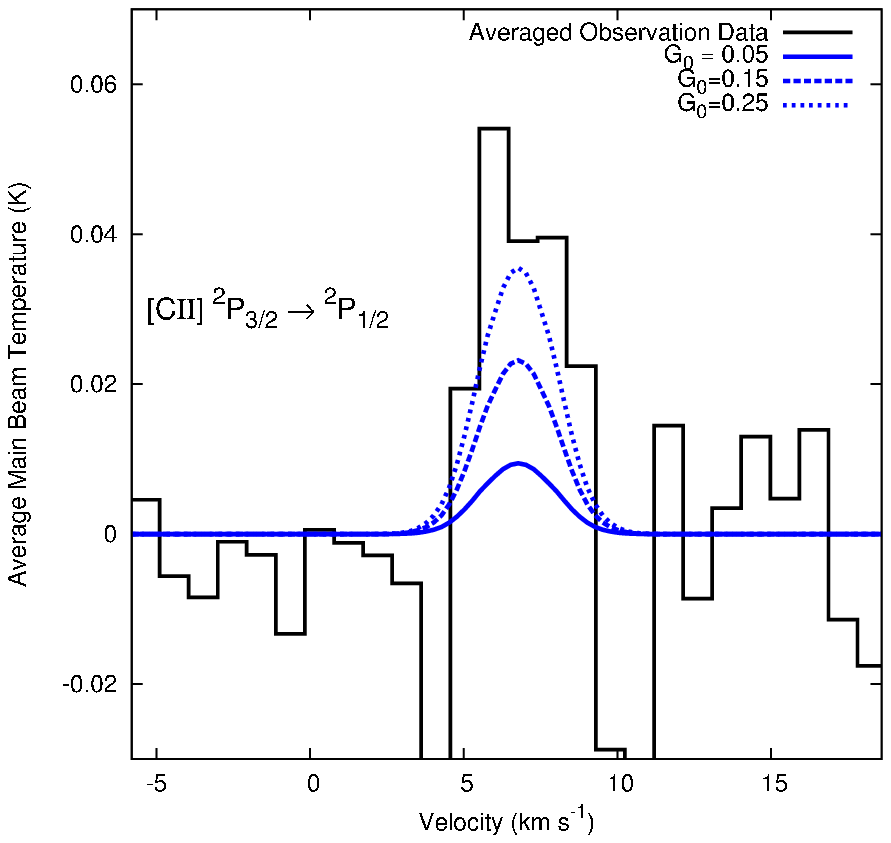}
\caption{Effect of variation of $G_0$ on the average [C\,{\sc ii}] 1900 GHz spectral intensity, as a function of velocity.  Models averaged over all positions compared with observational data for several values of $G_0$, with other parameters being: $\delta$(S) = 50, $^{12}$C/$^{13}$C = 43, and $\zeta=3\times10^{-17}$ s$^{-1}$.  Velocity axis of figure has been re-centered on the 6.78 km s$^{-1}$ velocity of the gaussian component of the averaged observations, and the modeled spectra are overlaid.  The rms uncertainty of the observation data is 0.04 K.  The model spectra assumed a doppler b-parameter of 1.6 km s$^{-1}$, in line with the averaged spectra's 2.64 km s$^{-1}$ FWHM.\\}
\label{fig:c+}
\end{figure}
\end{center}
\vspace{-1.2cm}

We were able to find general agreement between our model and observed integrated line intensities.  The model was able to reproduce the spatial distribution of the line-integrated intensities of $^{12}$CO and $^{13}$CO, across the linear edge region.  The line intensities were, however, over-predicted in our model by factors of up to 2, with the exception of $^{13}$CO, which was slightly under-predicted.  A typical fractional abundance and temperature profile generated by the Meudon code, as a function of depth into the cylindrical model, is found in Figure \ref{fig:Meudonout}.

In evaluating the results of our models, we quantified the closeness to which our models fit the observational data by employing a chi-squared measure of fit $\chi^2$ for each species, defined as
\begin{equation}
\chi^2=\frac{1}{N}\sum\limits_{i=1}^N \frac{(I_{model_i}-I_{obs_i})^2}{\sigma_{obs_i}^2}.
\end{equation}
Measuring $\chi^2$ allowed us to make distinctions between models across the broad parameter space.

In general, there is a reduction in [C\,{\sc i}], [C\,{\sc ii}], and $^{12}$CO line emission if a weaker FUV flux is assumed in a PDR because it reduces gas temperatures, and thereby the line excitation.  We explored the impact of varying the FUV field ($G_0 = 0.05 - 0.25$) on the [C	\,{\sc i}], [C\,{\sc ii}], $^{12}$CO, and $^{13}$CO line emission in our models.  Low values for $G_0$ are expected in the Taurus region \citep{pineda2010}.  For the $G_0$ values assumed, the models generally over-estimated the intensities of both  [C\,{\sc i}] lines (by as much as a factor of two), as well as that of the $^{12}$CO line emission.  On the other hand, the $^{13}$CO emission was too low, but differed by $< 30\%$ across most of the linear edge region.  The results of this variation in FUV field strength are shown in Figure \ref{fig:G0_var}.

It can be seen that the observational data are best represented by a PDR model with $G_0 = 0.05$, taking $\delta$(S)=50, $^{12}$C/$^{13}$C=43, and $\zeta=3\times10^{-17}$ s$^{-1}$.  For this model, we found $\chi^2_{\rm [C\,{\sc I}]_{10}}=91.4$, $\chi^2_{\rm [C\,{\sc I}]_{21}}=10.3$, $\chi^2_{\rm ^{12}CO}=29.1$, and $\chi^2_{\rm ^{13}CO}=10.9$ (where [C\,{\sc i}]$_{10}$ represents the $^3$P$_1\rightarrow^3$P$_0$ transition, and [C\,{\sc i}]$_{21}$ represents the $^3$P$_2\rightarrow^3$P$_1$ transition).  This FUV field is at the low end predicted by \citet{pineda2010} for the Taurus region, who found $G_0$ to be between 0.059 and 0.59.  The low external FUV field strength explains the weakness of the [C\,{\sc ii}] 1900 GHz line above rms noise levels at any single position, especially as this $G_0=0.05$ model predicts dust and gas temperatures well below 91.2 K, the excitation energy of the [C\,{\sc ii}] $^2$P$_{3/2}$ level above the $^2$P$_{1/2}$ ground state.

The external FUV field strength was the only parameter in our models whose variation affected [C\,{\sc ii}] emission.  Though we had marginal detections of [C\,{\sc ii}] 1900 GHz emission above the rms level at two positions, we detected a weak signal when the observations at all the positions were averaged.  We fit a gaussian to the averaged data and found a central velocity of 6.78 km s$^{-1}$, and a peak antenna temperature of 0.053 K with a FWHM of 2.64 km s$^{-1}$.  Similarly averaging the modeled [C\,{\sc ii}] spectra across the region, we compared the results of our best representative models, and variation in $G_0$, with this averaged spectra, as seen in Figure \ref{fig:c+}.  As expected of such a weak $G_0$, the modeled [C\,{\sc ii}] emission is extremely weak in the region, below the rms noise levels of the data ($\sigma \sim$ 0.04 K).  The averaged [C\,{\sc ii}] spectra suggest a higher $G_0$, inconsistent with that indicated by the data from other transitions. However, the [C\,{\sc ii}] emission may be explained by an unmodeled thin, hot and relatively diffuse foreground component.  We observed that the averaged spectral intensities were higher for the points observed outside the edge region (NE points), than those within it (SW points).  We also find that the model predicted higher intensities within the edge.  This inconsistency supports the presence of an additional unmodeled component.  Such a foreground cloud may also help to explain the hot molecular hydrogen component observed by \citet{goldsmith2010}, which requires $G_0 \gg 6$ to produce the necessary quantity of hot H$_2$ observed in what otherwise appears to be a rather quiescent region.

Variation of the initial isotopic ratio of carbon $^{12}$C/$^{13}$C almost singularly affected $^{13}$CO emission in our models.  In reducing the ratio, by increasing the initial abundance of $^{13}$C$^+$, more $^{13}$CO was indeed observed in the model, improving $\chi^2_{\rm ^{13}CO}$ from 37.9 for $^{12}$C/$^{13}$C = 77 to 10.9 for $^{12}$C/$^{13}$C = 43, with $G_0 = 0.05$, $\delta$(S) = 50, and $\zeta = 3 \times 10^{-17}$ s$^{-1}$.  Again, $^{13}$CO emission was still somewhat under-predicted in the best representative models.  The other species' integrated intensities were hardly affected by any isotopically-selective processes for any value of $^{12}$C/$^{13}$C used in our modeling, all changing in intensity by $<1\%$ across the region.  Though $^{13}$CO production is enhanced through isotopically-selective consumption of $^{12}$CO by a lower isotopic ratio \citep{langer1989}, the opacity of the $^{12}$CO transition in the models meant that the line intensity of this common isotopologue hardly changed as a result of this process.  Similarly, the C and C$^+$ column densities, and their subsequent line emission, were not greatly affected by the increased abundance of $^{13}$C$^+$.  It may be that values lower than $^{12}$C/$^{13}$C = 43 would produce $^{13}$CO integrated intensities which better fit observations, however such values are not supported by other work.
\begin{center}
\begin{figure}[!t]
\centering
\includegraphics[width=0.45\textwidth]{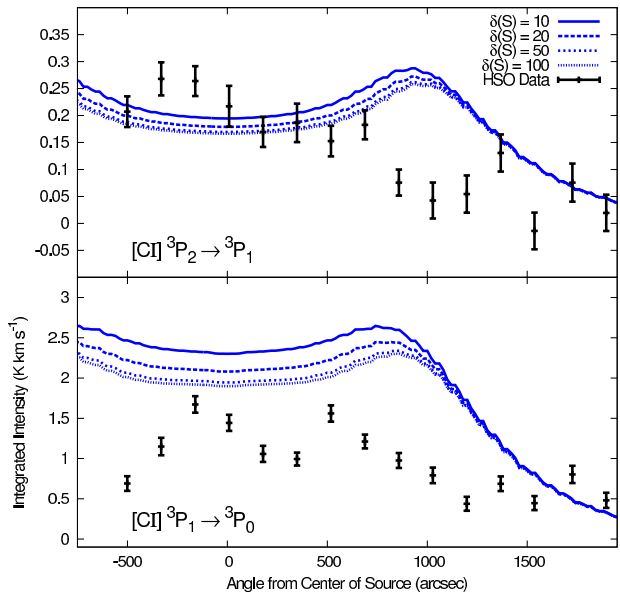}
\caption{Calculated integrated intensities of [C\,{\sc i}] transitions compared with observations for various factors of sulfur depletion $\delta$(S), across the cylindrical model, as a function of angular offset from its center.  For these models, $G_0=0.05$, $^{12}$C/$^{13}$C = 43, and $\zeta=3\times10^{-17}$ s$^{-1}$. Top panel: [C\,{\sc i}] $^3$P$_{2}\rightarrow^3$P$_{1}$  transition (809 GHz) model and data.  Bottom panel: [C\,{\sc i}] $^3$P$_1 \rightarrow ^3$P$_0$ transition (492 GHz) model and data.\\}
\label{fig:Sdep_CStacked}
\end{figure}
\end{center}
\vspace{-9mm}

The effects of depleting the initial sulfur abundance in the model, variation in $\delta$(S), were seen primarily in reduction of the [C\,{\sc i}] emission at 492 and 809 GHz.   The $\chi^2_{\rm [C\,{\sc I}]_{10}}$ fit of the [C\,{\sc i}] $^3$P$_1 \rightarrow ^3$P$_0$ transition was improved from 147.8 to 85.2 as $\delta$(S) was increased from 10 to 100.  The [C\,{\sc i}] $^3$P$_{2}\rightarrow^3$P$_{1}$ transition's $\chi^2_{\rm [C\,{\sc I}]_{21}}$ was less affected, being reduced from 12.3 to 10.1 by the same change of $\delta$(S).  The effects on the integrated intensity of the [C\,{\sc i}] lines for different values of $\delta$(S), while holding other parameters constant, can be seen in Figure \ref{fig:Sdep_CStacked}.  Diminishing effects on the line intensities are noted for values of $\delta$(S) $> 50$, whereupon the abundance of other metals becomes greater than sulfur.  The intensities of the observed transitions for $^{12}$CO, $^{13}$CO and [C\,{\sc ii}] were hardly affected by values of $\delta$(S) between 10 and 100, changing their integrated intensities by no more than 5\% at any point.  Though the electron number density is strongly dependent on the abundance of metals, and thus $\delta$(S), the model seems to indicate that cold H$_3^+$ is, as previous studies have found, not greatly affected by electron number density \citep{langer1989}, and in turn $\delta$(S) has little effect on the reaction pathways for $^{12}$CO and $^{13}$CO, demonstrated by their integrated intensities' near independence of $\delta$(S).

Our model did not quantitatively point to any particular value of $\delta$(S), but due to the coldness of the modeled region as indicated by reasonable agreement of our data with a low FUV field ($G_0 \sim 0.05$), $\delta$(S) may in fact be as large as 1000, as found in other cold molecular clouds by \citet{tieftrunk1994}.  Hotter regions are expected to exhibit lower $\delta$(S) values, as less sulfur would be adsorbed on the surface of dust grains in the molecular cloud, reducing depletion \citep{ruffle1999}. 
\begin{center}
\begin{figure}[!t]
\centering
\includegraphics[width=0.46\textwidth]{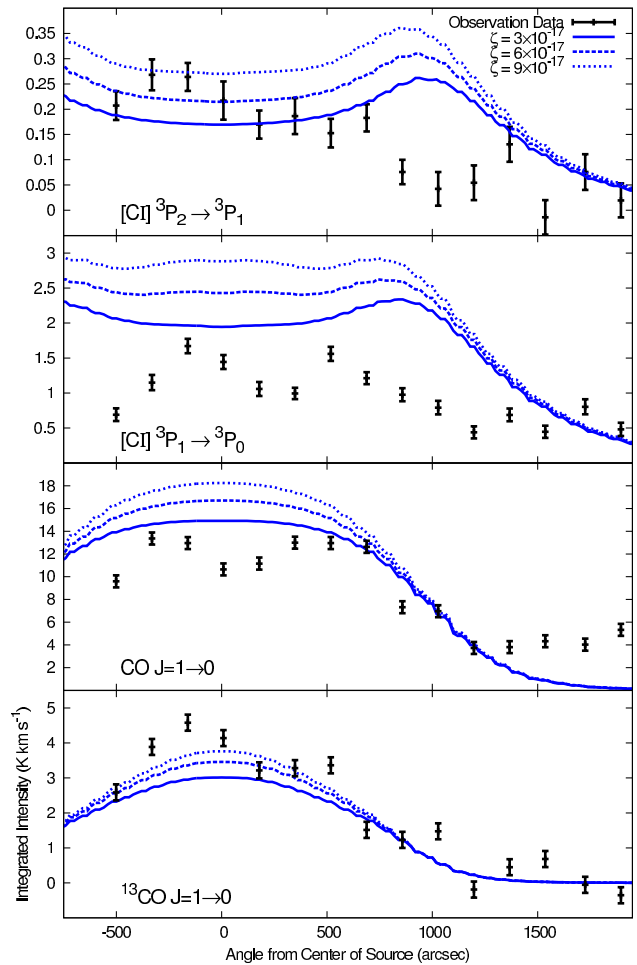}
\caption{Effect of variation of cosmic ray ionization rate $\zeta$ on [C\,{\sc i}], $^{12}$CO, and $^{13}$CO line-integrated intensities, as a function of angular offset from the center of the cylindrical model.  Models compared with observational data for several values of $\zeta$, with other parameters being: $G_0=0.05$, $\delta$(S) = 50, and $^{12}$C/$^{13}{\rm C} = 43$.  Panels from Top to Bottom:  [C\,{\sc i}] $^3$P$_{2}\rightarrow^3$P$_{1}$ (809 GHz),  [C\,{\sc i}] $^3$P$_1 \rightarrow ^3$P$_0$ (492 GHz), $^{12}$CO $J =1\rightarrow0$ (115 GHz), and $^{13}$CO $J =1\rightarrow0$ (110 GHz) transitions models and data.\\}
\label{fig:zeta_var}
\end{figure}
\end{center}
\vspace{-9mm}

Alternatively, dust grains themselves may be responsible for the effective reduction of [C\,{\sc i}] emission, rather than sulfur depletion, by depleting electron number density through both grain-assisted recombination and attenuation of carbon ionizing photons \citep{weingartner2001, li2001}.  As we do not have direct measures of dust grain size and density, nor sulfur abundance, we are unable to differentiate between the possible explanations for the reduced [C\,{\sc i}] emission.

Lastly, variation of the cosmic ray ionization rate, $\zeta$, significantly affected all observed species' integrated line intensities, with the exception of  [C\,{\sc ii}].  The integrated line intensities of $^{12}$CO, $^{13}$CO, and  [C\,{\sc i}] all increased with increasing $\zeta$.  Though higher values of $\zeta$ did help the model represent the $^{13}$CO data better (approximately halving $\chi^2_{\rm ^{13}CO}$ as $\zeta$ varied over the $(3-9)\times10^{-17}$ s$^{-1}$ range), the corresponding increases in the intensities of $^{12}$CO and [C\,{\sc i}] disimproved the overall fit with observations.  It was seen that the $\chi^2$'s for [C\,{\sc i}] and $^{12}$CO more than doubled from their values at $\zeta=3\times10^{-17}$ s$^{-1}$ as the ionization rate was increased to $9\times10^{-17}$ s$^{-1}$, while keeping $G_0=0.05$, $\delta$(S) = 50, and $^{12}$C/$^{13}$C = 43.

We found best agreement between the PDR model and observational data for a cosmic ray ionization rate $\zeta=3 \times 10^{-17}$ s$^{-1}$.  The effects of the variation of $\zeta$ can be clearly seen in Figure \ref{fig:zeta_var}.  Increased values for $\zeta$ had the effect of raising the temperature of dust and gas in the core of our cylindrical model.  For a depth of $A_{\rm V}$ = 1 mag., the gas temperature increased from 7.0 K for $\zeta=3\times10^{-17}$ s$^{-1}$ to 9.7 K for $\zeta=9\times10^{-17}$ s$^{-1}$, which primarily caused the increase in line intensity for [C\,{\sc i}], $^{12}$CO and $^{13}$CO.  These core temperatures are in good agreement with the 8--10 K temperatures of clouds determined in studies using $^{12}$CO \citep[e.g.][]{li2003}, and NH$_3$ \citep{tafalla2002}.  The increased core temperatures for higher values of $\zeta$ did not affect [C\,{\sc ii}] emission, as the heating occurs relatively deep into the cloud where C$^+$ is not abundant.

\subsection{Filament Inclination}\label{inc}
Our model implicitly assumes that the filament lies in the plane of the sky, perpendicular to the line of sight.  This, however, need not be the case.  In our modeling, we separately investigated the effects of inclination by scaling the central density of the filament as a function of angle $\theta$.  In order to remain consistent with observations, we held the column density of the model constant, accounting for the increased path length through the inclined cylinder by reducing the central density by a factor $\sec(\theta)$.  The Meudon outputs were then put into RATRAN as before, however the optically thin line emission calculated by RATRAN was scaled up by a factor $1/\sec(\theta)$ accounting for the longer path lengths.  The optically thick $^{12}$CO was not modified, though this introduced some error as an approximation.  For inclined models, we used the best-fit parameters of the non-inclined case, with $\delta$(S) = 50, $^{12}$C/$^{13}$C = 43, and $\zeta=3\times10^{-17}$ s$^{-1}$; the strength of FUV flux $G_0$ was varied from 0.05 to 0.25, as in the other models.  Investigating inclination angles between 0 and 75 degrees relative to the sky plane, we found that all angles increased the $\chi^2$, as the amount of material present was reduced significantly, in turn reducing line emission.  Increased values of $G_0$ were not able to reconcile both the magnitude and distribution of the reduced line emission in an appropriate manner.  Though we cannot rule out the existence of some small inclination, these models suggest that the filament does indeed lie close to the sky-plane, and that significant inclination is not supported by integrated emission data.

\subsection{Effects of Clumping}
A primary assumption made in our photochemical modeling was the choice of a radial density profile with a smoothly decreasing, quasi-power law profile for proton density (Eq.\,\ref{eq:king}), the parameters of which we fit to $A_{\rm V}$ data.  This smoothly varying model neglected the potential effects of clumping and small scale structures in the dust and gas.  Previous studies have used density inhomogeneities in the three-dimensional structure of their photochemical models to explain higher column densities of $^{12}$CO and $^{13}$CO relative to C and C$^+$ observed in the ISM as clumping results in a higher fraction of molecular dust and gas than otherwise expected \citep{stutzki1998}.

We investigated the effects of clumping in our model by means of ``fractional beam filling."  In order to simulate density inhomogeneities in the three-dimensional cloud with our one-dimensional density profile,  we scaled our density function by a scale factor ($f$), typically between 2 and 4, ran the model, and then divided the integrated intensities by the same factor.  This approximated a packing fraction for the gas, as the column density is directly proportional to the density profile.  Hence, scaling the density profile by a factor $f$ and dividing the intensity of the line transitions observed by the same factor is as though our antenna beam was only $1/f$ filled with dust and gas along the line of sight.  This approximated, to a rough degree, clumping and other density inhomogeneities in the linear edge region model.

In so doing, we observed a marked increase in the optical thickness of the $^{12}$CO and $^{13}$CO transitions, and that the peak temperature of the two [C\,{\sc i}] lines occurred at greater angular extents away from the projected center of the cloud, for models with 1/$f$ filling factors.  The fit was disimproved from the non-clumping case ($f=1$) for $^{12}$CO and $^{13}$CO, whose intensities were reduced overall, as temperatures within the denser cloud decreased to an extent not offset by the additional CO column density.  Additionally, although the $\chi^2$ for the 809 GHz [C\,{\sc i}] line was reduced, the bulk of the predicted [C\,{\sc i}] emission moved further out from the center of the observed region, and was much greater than observed.  We thus found that density inhomogeneities, on length scales much smaller than the characteristic scale of our modeling $a=0.457$ pc, were not necessary to explain the observed emission and that the presence of inhomogeneities was, in fact, not supported by our observations.

\subsection{Suprathermal Chemistry}
To produce additional $^{12}$CO and $^{13}$CO at low column densities, \citet{visser2009} suggested a suprathermal chemical pathway, due to Alfv\`en waves entering the PDR from outside, which results in non-thermal motions between ions and neutrals \citep[as suggested by][]{federman1996}, to produce additional $^{12}$CO and $^{13}$CO at low column densities ($A_{\rm V} \leq 0.3$ mag.).  As our model cylinder was only $\sim 2.6$ magnitudes thick through the center, we explored the effect of this suprathermal chemistry on the model by superimposing the column densities predicted by \citet{visser2009} for CO on the Meudon outputs for $A_{\rm V} < 0.1$ mag., where $N(^{12}$CO) and $N(^{13}$CO) predicted by \citet{visser2009} exceeded the Meudon outputs.  These were then input into RATRAN.  The line intensities for $^{12}$CO and $^{13}$CO were negligibly affected across the bulk of the region by the added CO at low $A_{\rm V}$, increasing by 0.12 K at most for $^{12}$CO and 0.02 K for $^{13}$CO anywhere, well under rms noise levels of the observations.  Our model filament already appears to be thick enough that conventional CO production processes account for the vast majority of observed $^{12}$CO and $^{13}$CO, especially as $\sim99.7\%$ of the $^{13}$CO occurs at visual extinction depths of $\geq0.2$ magnitudes in the PDR model (with $G_0=0.05$, $\delta$(S) = 50, $^{12}$C/$^{13}$C = 43, and $\zeta=3\times10^{-17}$ s$^{-1}$).

\section{Conclusions}\label{con}
In this paper we have presented observations of [C\,{\sc i}] and [C\,{\sc ii}] across the linear edge region in Taurus, as well as simulations of the region as a cylindrical PDR to model integrated line intensities of the [C\,{\sc i}] $^3$P$_1 \rightarrow ^3$P$_0$ (492 GHz) and $^3$P$_2 \rightarrow ^3$P$_1$ (809 GHz), [C\,{\sc ii}] $^2$P$_{3/2} \rightarrow ^2$P$_{1/2}$ (1900 GHz), and $^{12}$CO and $^{13}$CO $J =1 \rightarrow 0$ (115 and 110 GHz) transitions.  

We detected both  [C\,{\sc i}] lines in the linear edge region, where previous observations had found significant $^{12}$CO and $^{13}$CO emission.  However, except for possibly at two locations (NE4 and EDG), the [C\,{\sc ii}] 1900 GHz line was not detected above rms noise levels at any individual position.

We simulated the region as a cylindrical PDR using the Meudon PDR code and the RATRAN radiative transfer code, by taking the plane parallel output of the Meudon code and approximating it as the radial variation in the cylindrical model, which was then input to RATRAN.  Assuming the radius of curvature to be sufficiently large, we approximated a cut through the center of the spherical RATRAN model as a cut through a cylindrical model.  The sky-plane images generated by RATRAN were then analyzed to obtain line integrated intensities to compare with observations.  

The linear edge region's long and straight profile makes it an excellent case for employing a cylindrically symmetric density profile.  We fit the projected column density of our assumed cylindrical density function to the $A_{\rm V}$ data in \citet{pineda2010}.  We compared the model results with our  [C\,{\sc i}] and  [C\,{\sc ii}] integrated line intensities, as well as $^{12}$CO and $^{13}$CO $J = 1 \rightarrow 0$ integrated line intensity data from the linear edge presented in \citet{goldsmith2008}.

In our simulations, we tested the effects of small variations of the FUV field strength $G_0$, of changes in the isotopic abundance ratio of carbon $^{12}$C/$^{13}$C, of Sulfur depletion $\delta$(S), and of variations in the cosmic ray ionization rate $\zeta$.  Additionally, we considered the effects of clumping and suprathermal chemistry on our model.  We find that lower values of $G_0$ are required ($\sim 0.05$) in order that gas heating does not result in excessively high intensities of  [C\,{\sc i}] and $^{12}$CO transitions.  Values for $^{12}$C/$^{13}$C near the low end supported by literature ($\sim 43$) appear to best represent the linear edge region.  Sulfur depletion does not appear to play a large role in determining $^{12}$CO and $^{13}$CO emission, but we find evidence that this element must be significantly depleted ($\geq 50$) in order to explain the very low [C\,{\sc i}] emission.  An alternative to significant sulfur depletion may, however, be found in the reduction of electron number density due to dust grain-assisted recombination and attenuation of carbon-ionizing photons due to opacity contributions from grains \citep{weingartner2001, li2001}.

Additionally, we find that the cosmic ray ionization rate within the PDR, determined primarily by high-energy cosmic ray flux deep within the cloud ($A_{\rm V} \sim 1$ mag.), is likely on the order of $(3-6)\times10^{-17}$ s$^{-1}$, lying near the value predicted by \citet{goldsmith2005}.  This ionization rate reproduces dark core temperatures, on the order of 8--10 K, found by other studies of cold, relatively dense clouds \citep{tafalla2002,li2003}. Higher fluxes of low-energy cosmic rays, which could cause the cosmic ray ionization rate to be on the order of 10$^{-15}$ s$^{-1}$ in the outermost layers of the PDR, as proposed by \citet{mccall2003}, do not appear to be supported by the PDR model.  The modeled [C\,{\sc i}] line intensities were most sensitive to the cosmic ray ionization rate (see Figure \ref{fig:zeta_var}), and required low ionization rates and FUV fluxes in the outermost region of the PDR in order to agree with observations.

In addition to the PDR model using the Meudon code we investigated the effects of clumping and suprathermal chemistry on the PDR, as well as possible inclination of the filament relative to the sky-plane.  By introducing ``fractional beam filling", we roughly explored the effects of density inhomogeneities on the PDR.  Though it increased the $^{12}$CO and $^{13}$CO column density, temperatures within the clouds were lower, and thus reduced observed intensities, disimproving the overall fit of the models.  As well, the predicted [C\,{\sc i}] emission was greatly over-predicted with regard to observations.  Together, these effects indicate that density inhomogeneities on length scales smaller than the characteristic scale of our model $a=0.457$ pc are not supported by the observations.  We also roughly quantified the potential effects of suprathermal chemistry on $^{12}$CO and $^{13}$CO predicted by \citet{visser2009} by modifying the predicted Meudon output for $A_{\rm V} \leq 0.1$ mag.  Doing so had a negligible effect on $^{12}$CO and $^{13}$CO emission as it only added a small amount of material to the outer layers.  While suprathermal processes may be present, they do not appear necessary to explain our observations.  Investigating the possible effects of inclination of the filament relative to the sky-plane, we scaled the central density of the model as a function of inclination so as to keep the column density constant and in agreement with observations.  The resulting inclined model produced significantly less line emission for all species, significantly reducing the quality of fit.  Thus, significant inclination of the filament relative to the sky-plane appears to be contrary to our observations.


\section*{Acknowledgments}

We are grateful to W. D. Langer and M. Elitzur for useful discussions regarding photon-dominated regions and the astrochemistry associated with them.  We would also like to thank an anonymous referee for valuable suggestions, which significantly improved the paper.  This research was carried out at the Jet Propulsion Laboratory, which is operated by the California Institute of Technology, under contract by the National Aeronautical and Space Administration, as part of the JPL Summer Internship Program.  As well, this research made use of NASA's Astrophysics Data System, and of the Leiden Atomic and Molecular Database.

Facilities: \facility{HSO(HIFI)}, \facility{FCRAO}.

\bibliography{papers}
\bibliographystyle{apj.bst}

\end{document}